\documentclass[prl, reprint, amsmath,amssymb, aps]{revtex4-2}
\usepackage{graphicx}
\usepackage{dcolumn}
\usepackage{bm}
\usepackage{hyperref}
\usepackage{comment}
\hypersetup{colorlinks=true, linkcolor=blue, citecolor=blue, urlcolor=blue}

\begin{document}

\title{Robust Near-Critical Dynamics in Heavy-Tailed Neural Networks}

\author{Ryota Kojima}

\email{ryota_kojima_aa@mail.toyota.co.jp}

\affiliation{Toyota Motor Corporation, Toyota, Aichi, Japan}

\begin{abstract}
The criticality hypothesis posits that biological neural networks operate near a phase transition, yet within standard Gaussian mean-field theories this regime appears fragile and requires fine tuning. Here we show that heavy-tailed synaptic connectivity provides a robust alternative mechanism. By developing a dynamical mean-field theory for Cauchy-distributed couplings, we reduce the macroscopic dynamics to a one-dimensional gradient flow with a global Lyapunov potential. The resulting theory exhibits a continuous phase transition in which collective activity grows with the square root of the distance to criticality, and static susceptibility diverges only as the square root rather than linearly as in Gaussian mean-field theories. This structure gives rise to an emergent automatic gain control: activity-dependent noise fluctuations suppress the effective gain at high activity levels while preserving high susceptibility near the critical point. Extending this mechanism to general symmetric $\alpha$-stable inputs, we identify heavy-tailed synapses as a key microscopic origin of robust near-critical dynamics in disordered neural circuits.
\end{abstract}

\maketitle

\textit{Introduction.}---
The criticality hypothesis suggests that biological neural networks operate near a phase transition point to maximize computational capabilities such as dynamic range and information transmission \cite{Beggs2003, Bertschinger2004, Kinouchi2006, LANGTON199012, PhysRevE.84.051908, Munozarticle}. However, a fundamental theoretical tension exists between this hypothesis and standard mean-field theories based on Gaussian statistics \cite{PhysRevLett.61.259, doi:10.1126/science.274.5293.1724, Rajan2006, PhysRevX.8.041029}. In the Gaussian framework, criticality is typically associated with a sharp phase transition: the high-susceptibility regime is confined to a narrow vicinity of the critical coupling, making the system sensitive to parameter fluctuations and heterogeneity inherent in neural circuits\cite{Magnasco}.

Physiological evidence, by contrast, shows that synaptic weight distributions are highly skewed, often approximating log-normal or heavy-tailed statistics rather than Gaussian ones \cite{Sayer826, https://doi.org/10.1111/j.1469-7793.1999.00169.x, 10.1371/journal.pbio.0030068, Buzsaki2014, Cossell2015}. Such “strong-sparse’’ connectivity has been argued to facilitate efficient communication and information transfer \cite{teramae2012optimal, brunel2000dynamics}, and pioneering works have suggested that heavy-tailed connectivity can broaden the critical regime \cite{PhysRevLett.125.028101, Wardak2022}. However, these studies—as well as recent works focusing on modular or topological structures \cite{Kusmierz2025, Wainrib2016}—primarily relied on numerical simulations or approximate theories, reflecting the analytical intractability of heavy-tailed path integrals \cite{Cizeau1994, Feller1971}. Yet, despite being a plausible route toward reconciling the fragility of Gaussian models with the robustness of biological circuits, heavy-tailed connectivity still lacks a tractable theoretical framework.

In this Letter, we address this gap by constructing a dynamical mean-field theory (DMFT) for networks with heavy-tailed couplings. While recent DMFT formulations for generic non-Gaussian interactions lead to implicit integral relations \cite{Azaele2024, Pasqualini2026}, we focus on symmetric $\alpha$-stable synapses and demonstrate that for the Cauchy law ($\alpha = 1$), the DMFT closes to a one-dimensional macroscopic equation, analogous to low-dimensional reductions of recurrent circuits~\cite{Montbrio2015}. This maps the macroscopic dynamics to a gradient flow with a global Lyapunov potential, allowing us to derive the phase diagram in closed form. Crucially, it reveals a continuous phase transition with an order-parameter exponent $\beta = 1/2$ and a reduced susceptibility exponent $\gamma = 1/2$ (compared to the Gaussian $\gamma = 1$~\cite{Goldenfeld:1992qy}), confirming a broader non-Gaussian near-critical regime.

This gradient-flow structure naturally exposes an emergent physical mechanism that we term Automatic Gain Control (AGC). Activity-dependent fluctuations in the noise scale dynamically regulate the effective gain, providing a divisive-like reduction of macroscopic sensitivity as activity increases. This implements a form of activity-dependent gain control reminiscent of divisive normalization ~\cite{Carandini2012, Chance2002, Turrigiano2012}, but arizing directly from recurrent heavy-tailed synapses. For couplings near the critical point, AGC keeps the network responsive to weak inputs while automatically suppressing sensitivity at high activity levels. Thus, we identify heavy-tailed connectivity as a key microscopic origin of robust near-critical dynamics, offering a mathematically controlled alternative to standard Gaussian theories.

\begin{figure*}[t]
    \centering
    \begin{minipage}{0.32\textwidth}
        \centering
        \includegraphics[width=\linewidth]{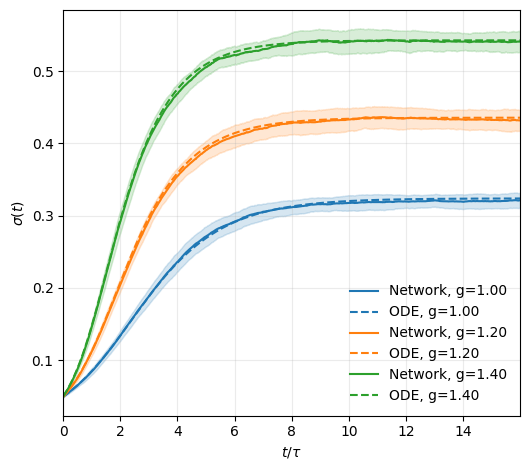}
    \end{minipage}
    \hfill 
    \begin{minipage}{0.66\textwidth}
        \centering
        \includegraphics[width=\linewidth]{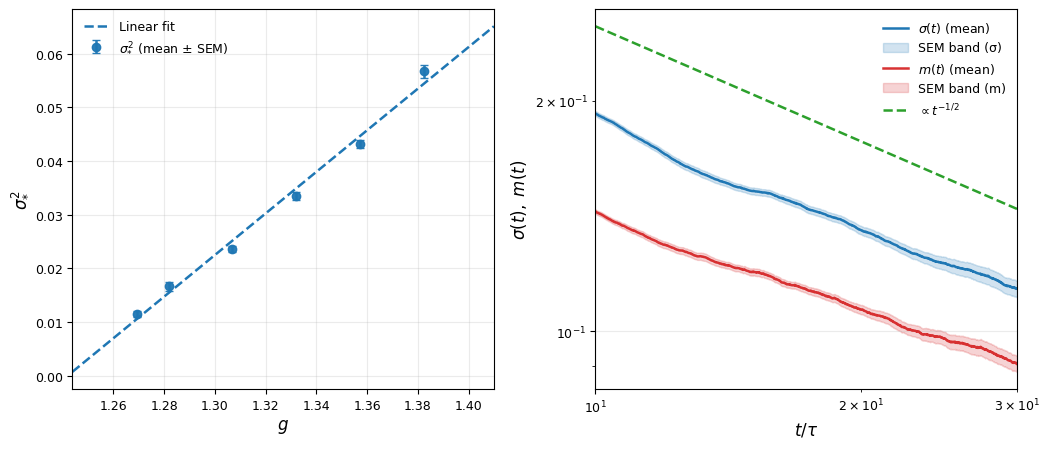}
    \end{minipage}
    
    \caption{Microscopic verification and continuous phase transition. 
    (Left) Solid curves show the macroscopic noise scale $\sigma(t)$ measured from microscopic network simulations. Here $\sigma$ is defined as the interquartile range/2 of the filtered input across neurons. Dashed curves show the solutions of the reduced one-dimensional ODE \eqref{eq:exact_ODE} initialized at the same $\sigma(0)$. 
    (Center) The squared stationary noise scale $\sigma_{*}^2$ (blue circles) scales linearly with the coupling strength $g$ near the onset of activity, consistent with the mean-field Landau exponent $\beta=1/2$. E
    (Right) Critical decay dynamics at $g \approx g_c$. Both the microscopic noise scale $\sigma(t)$ (blue) and the population activity $m(t)$ (red) exhibit algebraic decay $\propto t^{-1/2}$ (green dashed guide), consistent with the Landau mean-field universality class. }
    \label{fig:macro_and_critical}
\end{figure*}
\textit{Dynamical Mean-Field Closure}---We study large interacting networks with random, possibly heavy-tailed couplings and bounded rate outputs $\phi(x)$.
Each unit consists of a causal linear filter with impulse response $G$ (normalized so that $\|G\|_{L^1} = 1$), followed by a time-independent nonlinear output function $\phi$.
The input to unit $i$ is given by:
\begin{equation}
    u_{i}(t) := \sum_{j \ne i} J_{ij} \phi(x_{j}(t)) + h_{i}(t),
    \label{eq:input_def}
\end{equation}
where $h_{i}(t)$ is an external field. The bounded non-decreasing $\phi$ ensures statistical stability in the heavy-tailed regime and reflects typical neuronal gain curves.
The couplings $J_{ij}$ are drawn from an i.i.d. symmetric Cauchy distribution with scale parameter $g/N$, characterized by the characteristic function $\langle e^{ikJ_{ij}} \rangle = \exp[-\frac{g}{N}|k|]$.
Note that unlike the standard Gaussian scaling ($1/\sqrt{N}$), the $1/N$ scaling is required here to maintain an $O(1)$ input scale due to the heavy tails of the Cauchy distribution \cite{Azaele2024}.
The unit state is given by the causal convolution 
\begin{equation}
    x_{i}(t) = \int_{0}^{t} G(t-s) u_i(s)  ds\equiv (G * u_{i})(t).
\end{equation}
Here $x_{i}(t)$ denotes the filtered internal state (e.g. post-synaptic potential) of neuron $i$. 
Because the couplings are Cauchy-distributed, the endogenous input
$\eta_i(t)=\sum_j J_{ij}\phi(x_j)$ has Cauchy statistics at each fixed time.
We define the filtered input field $\tilde{x}(t) = (G * \eta)(t)$ and, averaging over the quenched disorder, obtain the characteristic functional
\begin{equation}
    \ln \langle e^{i \lambda \tilde{x}(t)} \rangle_{P[x]} = - g |\lambda| \left\langle \left| \int_{0}^{t} G(t-s) \phi(x(s)) ds \right| \right\rangle_{P[x]}.
    \label{eq:char_func_raw}
\end{equation}
where $P[x]$ is the effective single-site measure (see Supplemental Material \cite{SupplementalMaterial} for a detailed derivation).
For general kernels, the absolute value in \eqref{eq:char_func_raw} couples all times and prevents closure.
For nonnegative kernels ($G(t)\ge0$) and outputs ($\phi\ge0$), however, the convolution is nonnegative for any path, so the absolute value is redundant and Fubini's theorem allows us to commute the expectation and time integration:
\begin{equation}
    \ln \langle e^{i \lambda \tilde{x}(t)} \rangle_{P[x]} = - g |\lambda| \int_{0}^{t} G(t-s) \langle \phi(x(s)) \rangle_{P[x]} ds.
    \label{eq:char_func_closed}
\end{equation}
We recognize the right-hand side as the log-characteristic function of a Cauchy distribution with the time-dependent scale parameter
\begin{equation}
    \sigma(t) \equiv g (G * m)(t),
    \label{eq:sigma_closure}
\end{equation}
where $m(t) \equiv \langle \phi(x(t)) \rangle$ is the population firing rate.
Thus, at each time $t$, the filtered input $\tilde{x}(t)$ is symmetrically Cauchy-distributed with scale $\sigma(t)$.
In the presence of a spatially homogeneous drive $h(t)$, the filtered input decomposes as $\tilde{x}(t) = h(t) + \xi(t)$, where $\xi(t)$ is a symmetric Cauchy-distributed random variable with scale $\sigma(t)$.
Hence the firing rate is a function of the two scalar variables $m(t) = m(h(t),\sigma(t))$.

In this derivation we work under two assumptions: $G(t) \ge 0$ and a bounded nonnegative rate output $\phi(x) \ge 0$. 
These assumptions hold in many rate-based neuron models—for instance when $\phi$ encodes firing rates (e.g., Heaviside, sigmoids, saturated ReLUs) and $G(t)$ represents standard synaptic or dendritic kernels such as exponentials, alpha functions, or positive mixtures—and they simplify the closure by ensuring that the convolution is nonnegative for all activity histories. 
For sign-changing kernels the absolute value cannot be removed and a closure is no longer available; this case is treated below by deriving bounds via the triangle inequality.

In standard rate–based recurrent network models, the single–unit dynamics with a linear filter of time constant $\tau$ obey
\begin{equation}
    \tau \frac{dx_i(t)}{dt} = -x_i(t) + u_i(t),
\end{equation}
where $u_i(t)$ is the total input to unit $i$.
This corresponds to choosing the nonnegative exponential kernel $G(t) = \tau^{-1} e^{-t/\tau} \Theta(t)$, whose convolution $(G * u)(t)$ is the solution of the above equation.
Adopting this kernel in the closure \eqref{eq:sigma_closure} gives the closed macroscopic dynamics
\begin{equation}
    \tau \dot{\sigma}(t) = -\sigma(t) + g m(h(t), \sigma(t)).
    \label{eq:exact_ODE}
\end{equation}
This ODE represents a macroscopic description of the heavy-tailed network, reducing the infinite-dimensional stochastic problem to a single dynamical variable $\sigma(t)$. Microscopic network simulations confirm this analytical reduction [Fig.~\ref{fig:macro_and_critical}], showing that the population activity and noise scale follow the predicted one-dimensional dynamics.

\textit{Continuous Transition and Critical Slowing Down.}---
The nature of the phase transition is governed by the asymptotic behavior of the population activity $m(h,\sigma)$ near the quiescent state ($\sigma \to 0$).
For a Heaviside threshold $\phi(x) = \Theta(x - \theta)$, averaging over the Cauchy input yields
\begin{equation}
    m(0,\sigma) = \frac{1}{2} - \frac{1}{\pi} \arctan \left( \frac{\theta}{\sigma} \right) \ \mathop{\simeq}_{\sigma\to 0} \  \frac{\sigma}{\pi \theta} - \frac{\sigma^3}{3 \pi \theta^3}.
\end{equation}
A key feature of the heavy-tailed theory is that the linear term is nonzero, $m'(0)\neq 0$, in sharp contrast to Gaussian mean-field theories where Gaussian smoothing generically leads to $m'(0)=0$ and an onset of activity controlled by higher-order terms, typically via a saddle-node–like bifurcation.
Inserting this expansion into the macroscopic ODE~\eqref{eq:exact_ODE} yields, near the critical point,
\begin{equation}
    \tau\dot{\sigma} \simeq a \sigma - b \sigma^{3},
\end{equation}
i.e.\ a cubic Landau normal form with coefficients $a = g/(\pi\theta) - 1$ and $b = g/(3\pi\theta^{3})$.
The critical coupling $g_c=\pi\theta$, determined by the condition $a(g_c)=0$, agree with the critical point found in discrete-time Cauchy networks~\cite{PhysRevLett.125.028101}.
The system stabilizes at $\sigma_{*} = \sqrt{a/b}$ (for $a>0$), scaling as $\sigma_{*}\simeq \sqrt{3\theta/\pi}(g-g_{c})^{1/2}$.
This square-root scaling ($\beta = 1/2$) indicates a classical continuous second-order phase transition, as confirmed by the linear dependence of the squared order parameter $\sigma_*^2$ on $g$ shown in Fig.~\ref{fig:macro_and_critical}, and is also consistent with~\cite{PhysRevLett.125.028101}.
At criticality $(a=0)$, the cubic equation $\tau\dot{\sigma}=-b\sigma^{3}$ integrates to $\sigma(t) \sim t^{-1/2}$, demonstrating algebraic critical slowing down [Fig.~\ref{fig:macro_and_critical}].

Importantly, this mean-field universality class extends beyond the Heaviside model.
For a broad class of gain functions, the quadratic term in the small-$\sigma$ expansion is negligible in the far-subthreshold regime, so that the same cubic Landau normal form is recovered, with $\beta=1/2$ and $t^{-1/2}$ relaxation at criticality (see Supplemental Material~\cite{SupplementalMaterial}).
Thus, at the level of deterministic heavy-tailed DMFT, the continuous transition and the associated critical slowing down are robust features of Cauchy networks.

\begin{figure*}[t]
    \centering
    \begin{minipage}{0.66\textwidth}
        \centering
        \includegraphics[width=\linewidth]{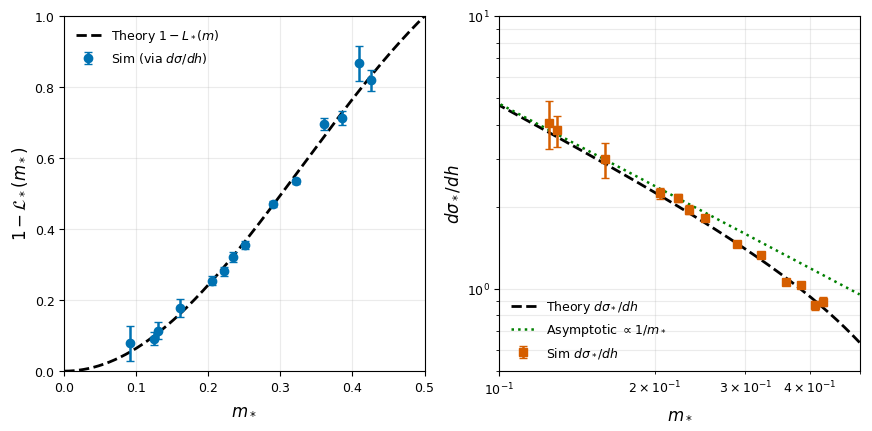}
    \end{minipage}
    \begin{minipage}{0.32\textwidth}
        \centering
        \includegraphics[width=\linewidth]{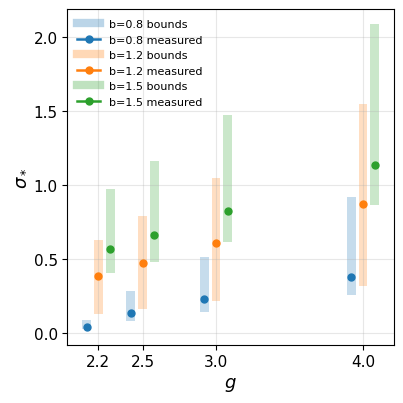}
    \end{minipage}
    \caption{(Left) Stability landscape. The curvature of the Lyapunov potential, $1 - \mathcal{L}_*$, is plotted against the stationary population rate $m_*$. The theoretical prediction (dashed line) is in agreement with numerical simulations (circles), indicating that the system remains in a stable regime over the explored activity range, while the curvature becomes small near criticality ($m_* \to 0$).
    (Center) Automatic Gain Control (AGC). The closed-loop susceptibility $d\sigma_*/dh$ scales inversely with activity. Simulation data (squares) follow the theoretical prediction (black dashed) and the asymptotic $1/m_*$ scaling (green dotted), illustrating the ``divisive brake". 
    (Right) Bounds for sign-changing kernels. We use a biphasic difference-of-exponentials kernel $G(t) = h_{\tau_1}(t) - b h_{\tau_2}(t)$ with $h_\tau(t)=\tau^{-1} e^{-t/\tau}\Theta(t)$, and various $b$. 
}
    \label{fig:agc_mechanism}
\end{figure*}

\textit{Near-Critical Dynamics and Automatic Gain Control.}---
We now characterize the dynamics of the noise scale $\sigma(t)$ near small-activity fixed points, using the gradient-flow structure of the Cauchy DMFT.
The macroscopic dynamics follow a gradient-flow structure governed by a global Lyapunov potential $\Phi(\sigma; h)$:
\begin{equation}
    \tau \dot{\sigma} = -\partial_\sigma \Phi(\sigma; h), \quad
    \Phi(\sigma; h) = \int_{0}^{\sigma} \big[s - g m(h,s)\big]  ds.
    \label{eq:potential}
\end{equation}
This potential landscape ensures that the noise scale $\sigma(t)$ relaxes monotonically toward a stable equilibrium $\sigma_*$, determined by the condition $\partial_\sigma \Phi = 0$ (which recovers the self-consistency equation $\sigma_* = g m(h, \sigma_*)$).
To assess stability, we evaluate the curvature of the potential at the fixed point, $\partial_\sigma^{2} \Phi|_{\sigma_*} = 1 - \mathcal{L}_*(h, \sigma)$, where
$\mathcal{L}_*(h, \sigma) \equiv g\partial_\sigma m(h,\sigma)\big|_{\sigma_*}$ represents the internal \textit{loop gain} of the noise feedback.
For the Heaviside step function, we derive the identity
\begin{equation}
    \mathcal{L}_*(h, \sigma) = \frac{\sin(2\pi m_*)}{2\pi m_*} \quad (< 1 \ \text{for} \ m_* > 0),
\end{equation}
where $m_* = m(h, \sigma_*)$. Since $\mathcal{L}_* < 1$ for any non-zero activity, the potential is strictly convex, guaranteeing that the stationary state is a unique attractor to which all trajectories converge, regardless of initial conditions [Fig.~\ref{fig:agc_mechanism}].
The curvature $1-\mathcal{L}_*$ also determines the speed of convergence, acting as a restoring force for the noise dynamics. 
Near the critical point ($m_* \to 0$, which for fixed input $h$ corresponds to $g$ approaching its critical value $g_c$), the loop gain approaches unity ($\mathcal{L}_* \to 1$), so that the curvature of the Lyapunov potential scales as $1-\mathcal{L}_* \propto m_*^2$. As a result, the linear relaxation time $\tau_{\mathrm{eff}} \sim \tau/(1-\mathcal{L}_*)$ diverges as $\tau_{\mathrm{eff}} \propto m_*^{-2}$, which is the hallmark of critical slowing down.
The functional consequences of this structure are revealed by the static response to a homogeneous input, quantified by the \textit{closed-loop susceptibility} $d\sigma_*/dh$.
Differentiating the stationary condition yields
\begin{equation}
    \frac{d\sigma_*}{dh} = \frac{g  A_{\mathrm{eff}}}{1-\mathcal{L}_*(h, \sigma)},
\end{equation}
where $A_{\mathrm{eff}}(h,\sigma) \equiv \partial_h m(h,\sigma)$ is the effective sensitivity of the population activity.
For the Heaviside case, the susceptibility can be computed explicitly and takes the form
\begin{equation}
\frac{d\sigma_*}{dh} = \frac{2\sin^2(\pi m_*)}{2\pi m_* - \sin(2\pi m_*)}.
\end{equation}
The susceptibility exhibits a dual character [Fig.~\ref{fig:agc_mechanism}]: near criticality ($m_* \to 0$), it diverges as $\propto 1/m_*$, providing high responsiveness to weak inputs, whereas in the high-activity limit ($m_* \to 1$) one finds $d\sigma_*/dh \to 0$, reflecting the strong suppression of sensitivity by self-generated noise.
We refer to this activity-dependent reduction of macroscopic sensitivity as a bidirectional \emph{Automatic Gain Control} (AGC) mechanism.
Functionally, it acts as a divisive-like brake: large self-generated noise $\sigma_*$ smooths the transfer function and effectively divides the gain, preventing runaway excitation while preserving high sensitivity at low activity.

From a critical-phenomena perspective, since $m_*(g)\propto (g-g_c)^{1/2}$ and $dm/d\sigma$ approaches a finite nonzero constant as $m_* \to 0$, the susceptibility of the mean activity,
\begin{equation}
  \chi(g)\equiv\left.\frac{d m_*}{d h}\right|_{h=0},
\end{equation}
inherits the same power-law divergence and scales as $\chi(g) \sim (g-g_c)^{-1/2}$, corresponding to a static critical exponent $\gamma=1/2$, in contrast to the Gaussian mean-field value $\gamma=1$~\cite{Goldenfeld:1992qy} (see Supplemental Material \cite{SupplementalMaterial} for details).
This weaker divergence implies that the near-critical high-susceptibility regime extends over a much broader range of coupling strengths than in Gaussian theories, i.e. heavy-tailed connectivity and AGC effectively widen the critical window.

A particularly transparent case arises when the threshold density follows a power law $\rho(\theta) \sim \theta^{\kappa-1}$ near the origin.
In this case, the loop gain and effective sensitivity become independent of the coupling strength $g$:
\begin{equation}
    \mathcal{L}_*(h, \sigma) = \kappa, \quad
    g A_{\mathrm{eff}} = \kappa\cot\left(\frac{\pi\kappa}{2}\right).
\end{equation}
Here, the dependence on $g$ cancels out: both the stability margin $1-\mathcal{L}_*(h, \sigma)$ and the effective sensitivity are fixed solely by the exponent $\kappa$.
For $\kappa$ close to one, this yields a near-marginal yet linearly stable regime that is robust to changes in synaptic gain. Thus, the combination of heavy-tailed connectivity and power-law threshold statistics provides a concrete example of how AGC can generate extremely $g$-robust near-critical behavior.

\textit{Extension to Symmetric $\alpha$-Stable Inputs and Sign-Changing Kernels.}---
The nonlinear closure above relies on two ingredients: Cauchy distribution ($\alpha=1$ within the family of symmetric $\alpha$-stable, $S\alpha S$, distributions) and a nonnegative kernel $G(t)\ge0$.
We now relax both assumptions and consider general $S\alpha S$ couplings with stability index $1<\alpha\le2$ and sign-changing kernels $G(t)$.
In this general setting, the absolute value in the characteristic functional can no longer be eliminated, and the infinite-dimensional dynamics do not reduce to the closed ODE of the Cauchy case.
Nevertheless, we can still derive kernel-agnostic bounds for the macroscopic noise scale.
From the characteristic functional of the filtered input under $S\alpha S$ couplings, we can identify an instantaneous scale $\sigma_\alpha(t)$ as
\begin{equation}
    \sigma_\alpha(t) := g^{\frac{1}{\alpha}}
    \left[\Big\langle\Big|\int_{0}^{t} G(t-s) \phi(x(s)) ds\Big|^\alpha\Big\rangle \right]^{\frac{1}{\alpha}}.
\end{equation}
In terms of this scale, we obtain two-sided bounds that depend only on coarse features of the kernel $G$ and the population activity $m(t)$.
For bounded, non-decreasing rates ($0\le \phi\le 1$), the stationary noise scale $\sigma_{\alpha*}$ satisfies
\begin{equation}
    g^{1/\alpha}|K_0| m_* \le \sigma_{\alpha*} \le g^{1/\alpha} m_*^{1/\alpha},
    \label{eq:alpha_dc}
\end{equation}
where $K_0=\int G$.
Equation~\eqref{eq:alpha_dc} shows that, irrespective of the sign structure of the synaptic kernel (i.e., even in the presence of inhibition), increasing the population activity $m_*$ inevitably leads to an expansion of the noise scale $\sigma_{\alpha*}$.
These bounds are confirmed by direct simulations with biphasic kernels (Fig.~\ref{fig:agc_mechanism}).
Since the input to each unit follows an $S\alpha S$ law, the effective sensitivity is bounded by the inverse of its scale, $A_{\mathrm{eff}} \lesssim C_\alpha / \sigma_\alpha$ where $C_\alpha$ is a constant depending on $\alpha$.
Combining this with the lower bound in \eqref{eq:alpha_dc}, we obtain a \emph{kernel-agnostic} upper bound on the effective sensitivity:
\begin{equation}
    A_{\mathrm{eff}}(h,\sigma_{\alpha*}) \lesssim \frac{C_\alpha}{g^{1/\alpha}|K_0|  m_*}.
    \label{eq:AGC_alpha}
\end{equation}
Full derivations are provided in the Supplemental Material \cite{SupplementalMaterial}. 

\textit{Conclusion}---
In summary, we have developed a non-Gaussian dynamical mean-field theory for recurrent neural networks with heavy-tailed synaptic connectivity. 
For networks with Cauchy-distributed couplings, the theory reduces to a solvable single-variable dynamics for the macroscopic noise scale, revealing a continuous transition with mean-field critical exponents and algebraic critical slowing down.
In this regime, activity-dependent fluctuations implement an automatic gain control: they dynamically suppress the effective sensitivity while preserving high susceptibility over a broad range of coupling strengths. 
By deriving kernel-agnostic bounds for general $\alpha$-stable synapses, including sign-changing kernels and balanced networks, we showed that this near-critical “safe” regime is robust to microscopic details of synaptic filtering and input statistics. Together, these results identify heavy-tailed synapses as a plausible mechanism for robust near-critical dynamics in disordered neural circuits.
They also suggest several directions for future work, including extending non-Gaussian mean-field theory to plastic and structured connectivity, linking the predicted activity-dependent gain control to in vivo measurements of cortical variability and responsiveness, and exploring how learning rules exploit heavy-tailed synapses to maintain near-critical operation.
\\ 

We thank Kazufumi Hosoda, Hiroyuki Suto, Hitoshi Yamada and Koji Kawahara for invaluable discussions and comments on the manuscript.
\bibliographystyle{apsrev4-2}
\bibliography{dmft}

\clearpage
\newpage
\onecolumngrid 
\setcounter{equation}{0}
\setcounter{figure}{0}
\setcounter{section}{0}
\renewcommand{\theequation}{S\arabic{equation}}
\renewcommand{\thefigure}{S\arabic{figure}}
\renewcommand{\thesection}{SM Sec. \Roman{section}}
\renewcommand{\thesubsection}{\Alph{subsection}}

\begin{center}
\textbf{\large Supplemental Material for: Robust Near-Critical Dynamics in Heavy-Tailed Neural Networks}
\end{center}

\section{Dimensionality Reduction for Cauchy Networks}
\label{sec:exact_reduction}

In this section, we derive the one-dimensional macroscopic closure for the Cauchy regime ($\alpha = 1$) with nonnegative kernels $G(t) \ge 0$, showing that the noise scale $\sigma(t)$ is given by a closed convolution equation in terms of the population activity $m(t)$. For exponential kernels this convolution reduces to a linear ODE, yielding an low-dimensional dynamical description.

\subsection{Characteristic Functional of the Filtered Input}
We consider the total input to the $i$-th unit,
$\eta_i(t) = \sum_{j\neq i} J_{ij} \phi(x_j(t))$, where $J_{ij}$ are i.i.d.\ symmetric $\alpha$-stable (S$\alpha$S) variables.
Since the probability density function $\rho(x)$ generally lacks a closed form, the distribution is defined by its characteristic function:
\begin{equation}
    \big\langle e^{ikJ_{ij}} \big\rangle
    \equiv \int_{-\infty}^{\infty} e^{ikx} \rho(x) dx
    = \exp\!\left[-\frac{g}{N} |k|^\alpha\right].
    \label{eq:sas_char_func}
\end{equation}
Thus the characteristic exponent is $g/N$, and the corresponding scale parameter of $J_{ij}$ is $\sigma_J = (g/N)^{1/\alpha}$, which ensures that the total input $\eta_i(t)$ remains of order $O(1)$ in the thermodynamic limit $N\to\infty$.

We then define the filtered input field as $\tilde{x}_i(t) = (G \ast \eta_i)(t) = \int_0^t G(t-s) \eta_i(s) ds$. The generalized central limit theorem ensures that, for each fixed time $t$, the distribution of $\eta_i(t)$ converges to an S$\alpha$S law, as verified for the Cauchy case ($\alpha=1$) by the Q–Q plot in Fig.~\ref{fig:microscopic_verification_cauchy}.

To characterize its statistics, we compute the generating functional $ \langle \exp(i \lambda \tilde{x}_i(t)) \rangle$.
Substituting the definition of $\eta_i$:
\begin{equation}
    \tilde{x}_i(t) = \sum_{j \neq i} J_{ij} \underbrace{\int_0^t G(t-s) \phi(x_j(s)) ds}_{U_j(t)}.
    \label{eq:input_field_def}
\end{equation}
Here, $U_j(t)$ is the filtered output of neuron $j$.
For fixed dynamical paths $\{x_j(\cdot)\}_{j\neq i}$, we now average over the quenched
couplings $\{J_{ij}\}$. Because the $J_{ij}$ are independent S$\alpha$S variables,
the characteristic function factorizes:
\begin{align}
    \big\langle e^{i \lambda \tilde{x}_i(t)} \big\rangle_{J}
    &= \prod_{j \neq i} \big\langle \exp\left( i \lambda J_{ij} U_j(t) \right) \big\rangle_{J} \nonumber \\
    &= \prod_{j \neq i} \exp\left( -\frac{g}{N} |\lambda U_j(t)|^\alpha \right) \nonumber \\
    &= \exp\left( -\frac{g}{N} |\lambda|^\alpha \sum_{j \neq i} |U_j(t)|^\alpha \right),
    \label{eq:char_func_factorized}
\end{align}
where $\langle \cdot \rangle_J$ denotes the average over the S$\alpha$S couplings $\{J_{ij}\}$.
Next, we average over the dynamical paths of the neurons $\{x_j\}$. In the thermodynamic
limit $N \to \infty$, the empirical average $\frac{1}{N}\sum_j |U_j(t)|^\alpha$
self-averages to the expectation $\langle |U(t)|^\alpha \rangle_{P[x]}$ with respect
to the effective single-site process $P[x]$. Thus, the log-characteristic function reads
\begin{equation}
    \ln \big\langle e^{i \lambda \tilde{x}(t)} \big\rangle_{P[x]}
    = -g |\lambda|^\alpha
      \left\langle \left| \int_0^t G(t-s) \phi(x(s))  ds \right|^\alpha \right\rangle_{P[x]},
    \label{eq:log_char_functional}
\end{equation}
where we have dropped the neuron index $i$ and written $\tilde{x}(t)$ for a
typical site.

\subsection{Commutativity of Path Averaging and Time Integration}
For the Cauchy case ($\alpha=1$), we establish the closure by commuting the path average $\langle \cdot \rangle_{P[x]}$ with the time integration, which relies on the Fubini–Tonelli theorem. In general, the absolute-value nonlinearity inside the path expectation prevents such a closure, but here it disappears thanks to the nonnegativity assumptions:
\begin{enumerate}
    \item The synaptic kernel is causal and nonnegative: $G(t) \ge 0$ for $t \ge 0$.
    \item The firing rate function is nonnegative: $\phi(x) \ge 0$.
\end{enumerate}
Under these conditions, the convolution integral is strictly non-negative for any path $x(s)$:
\begin{equation}
    \int_0^t G(t-s) \phi(x(s)) ds \ge 0.
    \label{eq:nonnegative_convolution}
\end{equation}
Therefore, the absolute value operation becomes the identity operator: $| \int \dots | = \int \dots$.
This allows us to commute the path averaging $\langle \cdot \rangle_{P[x]}$ with the time integration (Fubini's theorem):
\begin{align}
    \left\langle \left| \int_0^t G(t-s) \phi(x(s)) ds \right| \right\rangle_{P[x]} 
    &= \left\langle \int_0^t G(t-s) \phi(x(s)) ds \right\rangle_{P[x]} \nonumber \\
    &= \int_0^t G(t-s) \underbrace{\langle \phi(x(s)) \rangle_{P[x]}}_{m(s)} ds \nonumber \\
    &= (G \ast m)(t).
    \label{eq:commutation_result}
\end{align}
Substituting this back into the characteristic function:
\begin{equation}
    \ln \langle e^{i\lambda \tilde{x}(t)} \rangle_{P[x]} = -g |\lambda| (G \ast m)(t).
    \label{eq:log_char_cauchy}
\end{equation}
This form $-\sigma |\lambda|$ matches the characteristic function of a Cauchy distribution with scale parameter $\sigma$.
Thus, we have proven that the input $\tilde{x}(t)$ follows a Cauchy distribution with a time-dependent scale $\sigma(t)$ given by:
\begin{equation}
    \sigma(t) = g (G \ast m)(t).
    \label{eq:exact_closure_sigma_SM}
\end{equation}

For a standard exponential kernel $G(t) = \frac{1}{\tau}e^{-t/\tau}\Theta(t)$, $\sigma(t)$ is the solution to the linear ODE:
\begin{equation}
    \tau \frac{d\sigma(t)}{dt} = -\sigma(t) + g m(t).
    \label{eq:ode_sigma}
\end{equation}
This is the closed dynamical equation for the macroscopic scale $\sigma(t)$.
The population activity $m(t)$ is the expectation of $\phi(x)$ over the instantaneous Cauchy distribution $ \rho_t(x) = \frac{1}{\pi} \frac{\sigma(t)}{(x-h(t))^2 + \sigma(t)^2}$. For a Heaviside threshold $\phi(x) = \Theta(x-\theta)$:
\begin{equation}
    m(h, \sigma) = \int_\theta^\infty \frac{1}{\pi} \frac{\sigma}{(x-h)^2 + \sigma^2} dx = \frac{1}{2} - \frac{1}{\pi}\arctan\left(\frac{\theta-h}{\sigma}\right).
    \label{eq:heaviside_activity}
\end{equation}

\subsection{Internal loop gain for Heaviside thresholds}
\label{sec:loop_gain}

For the Heaviside threshold $\phi(x)=\Theta(x-\theta)$, the closed Cauchy DMFT allows us to compute the internal loop gain at the fixed point defined as
\begin{equation}
    \mathcal{L}_*(h, \sigma) \equiv g  \partial_\sigma m(h, \sigma)\big|_{\sigma_*},
    \label{eq:def_L}
\end{equation}
and $m(h, \sigma)$ is given by \eqref{eq:heaviside_activity}.
Differentiating with respect to $\sigma$ yields
\begin{equation}
    \frac{\partial m}{\partial \sigma} = \frac{1}{\pi\sigma} \frac{\xi}{1+\xi^2}, \quad \text{with } \xi \equiv \frac{\theta-h}{\sigma}.
    \label{eq:diff_m_sigma}
\end{equation}
We analyze stability on the stationary manifold defined by the self-consistency condition $\sigma_* = g  m(h, \sigma_*)$ (assuming $\|G\|_1=1$).
Substituting $g = \sigma_*/m(h, \sigma_*)$ into the loop gain eliminates $g$ and $\sigma_*$:
\begin{equation}
    \mathcal{L}_*(h, \sigma)
    = \frac{\sigma_*}{m_*} \frac{1}{\pi\sigma_*} \frac{\xi_*}{1+\xi_*^2}
    = \frac{1}{\pi m_*} \frac{\xi_*}{1+\xi_*^2}.
    \label{eq:loop_gain_u}
\end{equation}
where $\xi_*=(\theta-h)/\sigma_*$ and $m_* = m(h, \sigma_*)$. To eliminate the parameter $\xi_*$, we invert the rate equation
\begin{equation}
    m_* = \frac{1}{2} - \frac{1}{\pi}\arctan(\xi_*).
\end{equation}
Using $\tan(\tfrac{\pi}{2} - x) = \cot(x)$, we obtain
\begin{equation}
    \xi_* = \tan\Big(\frac{\pi}{2} - \pi m_*\Big) = \cot(\pi m_*).
    \label{eq:u_y_relation}
\end{equation}
Substituting into \eqref{eq:loop_gain_u} and using $1+\cot^2\phi = \csc^2\phi$,
\begin{equation}
    \frac{\xi_*}{1+\xi_*^2} = \frac{\cot(\pi m_*)}{\csc^2(\pi m_*)}
    = \frac{1}{2}\sin(2\pi m_*),
    \label{eq:trig_identity}
\end{equation}
so that
\begin{equation}
    \mathcal{L}_*(h, \sigma) = \frac{\sin(2\pi m_*)}{2\pi m_*}.
    \label{eq:loop_gain_sinc}
\end{equation}
Since $\sin(x) < x$ for any $x > 0$, it follows that $\mathcal{L}_* < 1$ pointwise for any non-zero activity $m_* > 0$.
As $m_* \to 0$, we have $\mathcal{L}_* \to 1$, which matches the smooth approach to criticality described by the Landau expansion in the main text.

\subsection{Critical scaling of the susceptibility}
\label{sec:critical_susceptibility}
Using the Cauchy DMFT and the loop gain derived above, we now compute the critical scaling of the static susceptibility
 $\chi(g) \equiv \left.d m_*/d h\right|_{h=0}$  for the Heaviside case.

We first recall the small-$\sigma$ expansion of the rate function
(cf.\ Eq.~\eqref{eq:heaviside_activity}) at $h=0$:
\begin{equation}
  m(0,\sigma)
  = \frac{1}{2}
    -\frac{1}{\pi}\arctan \left(\frac{\theta}{\sigma}\right)
  = \frac{\sigma}{\pi\theta}
    -\frac{\sigma^3}{3\pi\theta^3}
    +O(\sigma^5).
  \label{eq:m_small_sigma_appendix}
\end{equation}
Combining this with the self-consistency condition
$\sigma_* = g m(0,\sigma_*)$ (Eq.~\eqref{eq:exact_closure_sigma_SM}), one finds
for $g$ slightly above the critical coupling $g_c=\pi\theta$:
\begin{equation}
  \sigma_*(g)\propto\sqrt{g-g_c},
  \quad
  m_*(g) = m(0,\sigma_*) \propto\sqrt{g-g_c},
  \label{eq:mstar_scaling_appendix}
\end{equation}
corresponding to the order-parameter exponent $\beta=1/2$ as stated in the main text.
The internal loop gain on the stationary manifold is given by
Eq.~\eqref{eq:loop_gain_sinc}. Expanding for small $m_*$ yields
\begin{equation}
  \mathcal{L}_*(h, \sigma) = 1 - \frac{2\pi^2}{3}m_*^2 + O(m_*^4),
\end{equation}
so that
\begin{equation}
  1-\mathcal{L}_*(h, \sigma)\propto m_*^2\propto(g-g_c),
  \label{eq:one_minus_L_scaling}
\end{equation}
and the relaxation time $\tau_{\mathrm{eff}}=\tau/(1-\mathcal{L}_*)$ diverges as $\tau_{\mathrm{eff}}\propto (g-g_c)^{-1}$, reflecting dynamical critical slowing down.

The closed-loop susceptibility of the noise scale is obtained by
differentiating the stationary condition $\sigma_*=g m(h,\sigma_*)$ with
respect to the homogeneous input $h$:
\begin{equation}
  \frac{d\sigma_*}{dh}
  = \frac{g A_{\mathrm{eff}}}{1-\mathcal{L}_*(h, \sigma)},\quad
  A_{\mathrm{eff}}(h,\sigma) \equiv \partial_h m(h,\sigma),
  \label{eq:dsigma_dh_appendix}
\end{equation}
in agreement with the expression used in the main text.
For the Heaviside gain, differentiating Eq.~\eqref{eq:heaviside_activity}
with respect to $h$ gives
\begin{equation}
  A_{\mathrm{eff}}(h,\sigma) = \frac{1}{\pi}\frac{1}{1+((\theta-h)/\sigma)^2}\frac{1}{\sigma}.
\end{equation}
At $h=0$ and for $\sigma\ll\theta$, this reduces to
\begin{equation}
  A_{\mathrm{eff}}(0,\sigma)
  \simeq \frac{1}{\pi}\frac{\sigma^2}{\theta^2}\frac{1}{\sigma} = \frac{\sigma}{\pi\theta^2}.
\end{equation}
Evaluated at the stationary value $\sigma=\sigma_*(g)$, and using
\eqref{eq:mstar_scaling_appendix}, we obtain
\begin{equation}
  A_{\mathrm{eff}}(0,\sigma_*)\propto \sqrt{g-g_c}.
  \label{eq:Aeff_scaling}
\end{equation}
Combining Eqs.~\eqref{eq:one_minus_L_scaling},
\eqref{eq:dsigma_dh_appendix}, and \eqref{eq:Aeff_scaling}, we find
\begin{equation}
  \frac{d\sigma_*}{dh}\propto \frac{\sqrt{g-g_c}}{g-g_c}\propto (g-g_c)^{-1/2}.
  \label{eq:dsigma_dh_scaling}
\end{equation}
The susceptibility of the mean activity is then
\begin{equation}
  \chi(g)\equiv\left.\frac{d m_*}{d h}\right|_{h=0}
  = \left.\frac{\partial m}{\partial \sigma}\right|_{\sigma_*}\frac{d\sigma_*}{dh} + A_{\mathrm{eff}}(0,\sigma_*).
\end{equation}
From Eq.~\eqref{eq:m_small_sigma_appendix}, we have
 $\partial m/\partial\sigma \to 1/(\pi\theta)$  as $\sigma\to 0$, i.e.\ a
finite, nonzero constant, whereas $A_{\mathrm{eff}}(0,\sigma_*)$ vanishes as $\propto\sqrt{g-g_c}$ [Eq.~\eqref{eq:Aeff_scaling}] and is therefore subleading compared to the first term in the vicinity of the critical point.
Using \eqref{eq:dsigma_dh_scaling}, we conclude that
\begin{equation}
  \chi(g) \propto (g-g_c)^{-1/2}.
\end{equation}
Thus the static susceptibility exponent for the Heaviside Cauchy network is
\begin{equation}
  \gamma = \frac{1}{2},
\end{equation}
in contrast to the Gaussian mean-field value $\gamma=1$. 

\begin{figure*}[t]
    \centering
    \includegraphics[width=\textwidth]{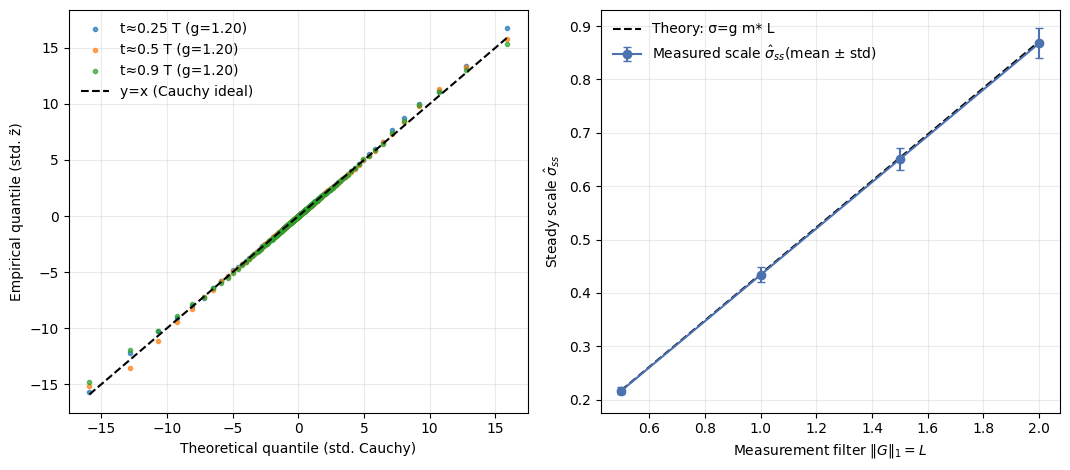} 
    \caption{(Left) Quantile–Quantile (Q–Q) plot of the effective input field $\tilde{x}(t)$ filtered by an exponential kernel $G(t) = \tau^{-1} e^{-t/\tau}\Theta(t)$ with $\tau=1$ (corresponding to $L=1$), against a standard Cauchy distribution. The data collapsing onto the diagonal $y=x$ confirms that the inputs follow the Cauchy statistics predicted by the mean-field theory.
    (Right) Validation of the $L^1$-norm scaling law using a family of exponential measurement filters $G_L(t) = (L/\tau)  e^{-t/\tau}\Theta(t)$ with $\tau=1$ and $L\in\{0.5,1.0,1.5,2.0\}$, whose $L^1$ norm is $\|G_L\|_1=L$. The stationary noise scale $\sigma_*$ estimated from the filtered trajectories scales linearly with $L=\|G_L\|_1$, in agreement with the theoretical prediction $\sigma_* = g m_* L$. Error bars indicate the standard deviation across 20 disorder realizations.}
    \label{fig:microscopic_verification_cauchy}
\end{figure*}

\section{L$^{1}$-Norm Scaling of Filtered Cauchy Input and Non-Self-Averaging}
\label{sec:L1_scaling_SM}
In this section we derive the $L^{1}$-norm scaling law for filtered Cauchy input fields and relate it to non-self-averaging properties of time-averaged observables.
\subsection{L$^{1}$ scaling for nonnegative causal test filters}
For the Cauchy case ($\alpha=1$) with nonnegative synaptic kernel $G(t)\ge 0$ and
nonnegative rate function $\phi(x)\ge 0$, we can use the DMFT closure derived above
(see Eq.~\eqref{eq:exact_closure_sigma_SM}).
In a stationary regime with constant activity $m(t)\equiv m_*$, the scale $\sigma(t)$ is constant in time, and the input field $\eta(t)$ is strictly stationary.
To probe how Cauchy noise propagates through causal linear filters, consider an auxiliary output
\begin{equation}
    Y_f \equiv \int_{0}^{\infty} f(t) \eta(t) dt,
\end{equation}
where $f\in L^1(\mathbb{R}_+)$ is a nonnegative causal test filter, $f(t)\ge 0$.
The corresponding characteristic function is obtained by inserting the probe function $\lambda f(t)$ into the characteristic functional of $\eta$.
Repeating the steps leading to Eq.~\eqref{eq:log_char_functional} with $G$ replaced by $f$ and using stationarity, we obtain
\begin{equation}
    \ln \left\langle e^{i \lambda Y_f} \right\rangle
    = - g |\lambda|\left\langle \int_{0}^{\infty} f(t) \phi(x(t)) dt \right\rangle
    = - g |\lambda|   \langle\phi\rangle \int_{0}^{\infty} f(t) dt,
    \label{eq:L1_scaling_SM}
\end{equation}
where $\langle\phi\rangle = m_*$ is the stationary mean firing rate.
For nonnegative $f$, its $L^1$ norm reduces to
\begin{equation}
    \|f\|_{L^1} := \int_{0}^{\infty} |f(t)| dt = \int_{0}^{\infty} f(t) dt.
\end{equation}
Equation~\eqref{eq:L1_scaling_SM} therefore shows that, in the Cauchy regime, the scale of the filtered observable $Y_f$ depends only on the $L^1$ norm (area) of the filter, in sharp contrast to Gaussian theories, where the variance scales with the $L^2$ norm.

\subsection{Implications for non-self-averaging}
The $L^1$-norm scaling reflects the well-known fact that linear functionals of Cauchy-distributed noise remain Cauchy, with a scale proportional to the integral of the absolute kernel.
A simple and important consequence is strong non-self-averaging: time averages of Cauchy observables do not concentrate around ensemble means as the observation window grows.
For instance, if one chooses $f_T(t) = T^{-1} \mathbf{1}_{[0,T]}(t)$, then $Y_{f_T}$ is the temporal average of $\eta(t)$ over a window of length $T$, and Eq.~\eqref{eq:L1_scaling_SM} yields
\begin{equation}
    \ln \left\langle e^{i \lambda Y_{f_T}} \right\rangle
    = - g \langle\phi\rangle |\lambda| 
      \|f_T\|_{L^1}
    = - g \langle\phi\rangle |\lambda|,
\end{equation}
independent of $T$.
Thus, the distribution of the time average remains a Cauchy law with \emph{finite} width even as $T\to\infty$, and time averages fail to converge to a single deterministic value.
This illustrates the strong non-self-averaging of Cauchy networks: macroscopic observables can exhibit persistent sample-to-sample fluctuations despite long temporal averaging.

Microscopic simulations in Fig.~\ref{fig:microscopic_verification_cauchy} confirm this $L^1$-scaling, illustrating the non-self-averaging character of Cauchy inputs.

\section{Universality Beyond Heaviside: General Nonnegative Rate Functions}
\label{sec:general_J}
In this section we demonstrate that the core results—the closure, the Landau–Ginzburg effective dynamics with exponent $\beta=1/2$, and the associated critical slowing down—extend to a broad class of bounded, non-decreasing rate functions $\phi(x)$.
This confirms that the near-critical ``safe'' regime generated by heavy-tailed connectivity is a robust feature, independent of the microscopic details of the neuronal transfer function.
The population activity $m(h, \sigma)$ is given by the convolution of the bare gain function $\phi(x)$ with the Cauchy kernel.
This becomes particularly transparent in the Fourier domain.
Using the convolution theorem, the spatial integral turns into a product, where the Cauchy kernel transforms into the exponential amplitude $e^{-\sigma |k|}$.
Identifying the wavenumber factor $|k|$ with the fractional Laplacian operator $|\partial|$ upon inverse transformation, we obtain
\begin{equation}
\begin{aligned}
    m(h,\sigma) 
    &= \int_{-\infty}^{\infty}\frac{1}{\pi}\frac{\sigma}{(x-h)^2+\sigma^2}  \phi(x)  dx \\
    &= \mathcal{F}^{-1} \left[ e^{-\sigma |k|} \widehat{\phi}(k) \right](h) \\
    &= e^{-\sigma |\partial|} \phi(h),
\end{aligned}
\label{eq:poisson_semigroup}
\end{equation}
where $\mathcal{F}^{-1}$ denotes the inverse Fourier transform and $\widehat{\phi}(k)$ is the Fourier transform of $\phi$.
This representation highlights that the scale parameter $\sigma$ acts as an effective ``diffusion time'', smoothing the transfer function via the fractional diffusion operator.
\subsection{Landau expansion and critical exponents}
To analyze the phase transition, we consider the small-$\sigma$ expansion of the smoothing operator.
Since $|\partial|$ is non-local, we use the identity $|\partial| = \partial_x \mathcal{H}$, where $\mathcal{H}$ is the Hilbert transform defined as a principal-value convolution with a translation-invariant kernel.
Writing $h$ for the spatial variable, we define
\begin{equation}
  (\mathcal{H}f)(h)
  = \frac{1}{\pi} \mathrm{p.v.}
      \int_{-\infty}^{\infty}\frac{f(x)}{h-x} dx
  = (K*f)(h),
\end{equation}
where the convolution kernel depends only on the difference $h-x$,
\begin{equation}
  K(y) = \frac{1}{\pi} \mathrm{p.v.}\frac{1}{y},\qquad y = h-x.
\end{equation}
In Fourier space, the Hilbert transform acts as a simple multiplier.
Denoting the Fourier transform by $\widehat{f}(k)=\int_{-\infty}^{\infty}e^{-ikx}f(x) dx$, the convolution theorem gives
\begin{equation}
  \widehat{\mathcal{H}f}(k)
  = \widehat{K*f}(k)
  = \widehat{K}(k) \widehat{f}(k).
\end{equation}
Using the standard identity $\widehat{\mathrm{p.v.}(1/x)}(k)=-i\pi \mathrm{sgn}(k)$, we obtain
\begin{equation}
  \widehat{K}(k)
  = \frac{1}{\pi} \widehat{\mathrm{p.v.}(1/y)}(k)
  = -i \mathrm{sgn}(k),
\end{equation}
and therefore
\begin{equation}
  \widehat{\mathcal{H}f}(k)
  = -i \mathrm{sgn}(k) \widehat{f}(k).
\end{equation}
In other words, $\mathcal{H}$ corresponds to the Fourier multiplier $-i \mathrm{sgn}(k)$.
Therefore $\partial_x\mathcal{H}$ has symbol $(ik)(-i \mathrm{sgn}(k)) = |k|$, matching the spectrum of $|\partial|$.
Expanding the smoothing operator for small $\sigma$ gives
\begin{equation}
\begin{aligned}
    m(h,\sigma) &= e^{-\sigma \partial_x \mathcal{H}}  \phi(h) \\
    &= \left( 1 - \sigma \partial_x \mathcal{H} 
              + \frac{\sigma^2}{2}(\partial_x \mathcal{H})^2 
              - \frac{\sigma^3}{6} (\partial_x \mathcal{H})^3 + \dots \right) \phi(h) \\
    &= \phi(h) - \sigma  \mathcal{H}(\phi')(h) 
       + \frac{\sigma^2}{2}  \phi''(h) 
       + \frac{\sigma^3}{6}  \mathcal{H}(\phi''')(h) + O(\sigma^4),
\end{aligned}
\label{eq:landau_expansion}
\end{equation}
where we used $(\partial_x \mathcal{H})\phi = \mathcal{H}(\phi')$ and $(\partial_x\mathcal{H})^2=-\partial_x^2$.
The first identity follows from the translation-invariance of the Hilbert kernel.
The identity $\partial_h[1/(h-x)] = -\partial_x[1/(h-x)]$ allows differentiation 
to act on $\phi$:
\begin{equation}
\partial_h (\mathcal{H}\phi)(h) 
= \frac{1}{\pi} \mathrm{p.v.} \int_{-\infty}^\infty \frac{\phi'(x)}{h-x}  dx
= (\mathcal{H}\phi')(h) .
\end{equation}
The second identity follows by applying 
this result twice:
\begin{equation}
(\partial_x\mathcal{H})^2 \phi = \mathcal{H}(\mathcal{H}(\phi'')) = -\phi'' ,
\end{equation}
using $\mathcal{H}^2 = -I$.
Matching powers of $\sigma$ with the Landau expansion
\begin{equation}
 m(h,\sigma)\simeq \phi(h) + c_1(h)\sigma + c_2(h)\sigma^2 + c_3(h)\sigma^3 + \dots
\end{equation}
we identify
\begin{equation}
    c_1(h) = \partial_\sigma m|_{\sigma=0} = -\mathcal{H}(\phi')(h),\quad
    c_2(h) = \frac12  \phi''(h),\quad
    c_3(h) = \frac16 \partial_\sigma^3 m|_{\sigma=0}
           = \frac16 \partial_x^2\mathcal{H}(\phi')(h).
\label{eq:landau_coeffs_all}
\end{equation}
Here $c_1$ and $c_3$ involve the \emph{non-local} Hilbert transform of $\phi'$, whereas $c_2$ is purely \emph{local}.
This structural difference leads to distinct decay laws in the far-subthreshold regime $\Delta:=\theta-h\gg 1$, where $\theta$ is the effective threshold of the gain curve.
\paragraph*{Nonlocal $c_1$ and $c_3$.}
The $1/(h-x)$ kernel implies long–range contributions from the threshold region.
Writing $x = \theta+v$ and expanding the kernel for $h$ well below threshold,
\begin{equation}
\frac{1}{h-x}
= \frac{1}{h-\theta-v}
= -\frac{1}{\Delta}\sum_{k=0}^\infty\Big(-\frac{v}{\Delta}\Big)^k
= \sum_{k=0}^\infty(-1)^{k+1}\frac{v^k}{\Delta^{k+1}},\quad \Delta:=\theta-h,
\end{equation}
and substituting into $c_1=\mathcal{H}(\phi')$ yields
\begin{equation}
c_1(h)=\frac{1}{\pi}\sum_{k=0}^\infty(-1)^k\frac{\mu_k}{\Delta^{k+1}},
\label{eq:c1expansion}
\end{equation}
with multipole moments $\mu_k=\int_{-\infty}^{\infty} v^k \phi'(v+\theta) dv$.
Differentiating twice with respect to $h$ gives
\begin{equation}
c_3(h)=\frac{1}{6\pi}\sum_{k=0}^\infty\frac{(-1)^k(k+1)(k+2) \mu_k}{\Delta^{k+3}}.
\end{equation}
The leading monopole term $\mu_0>0$ (for any non-decreasing $\phi$) gives algebraic decay:
\begin{equation}
c_1(h)\sim \frac{\mu_0}{\pi\Delta},\quad c_3(h)\sim \frac{\mu_0}{3\pi\Delta^{3}}\quad (\Delta\to\infty).
\end{equation}
These scalings are independent of whether $\phi$ is threshold-like or smooth.
\paragraph*{Local $c_2$.}
We next examine the decay rate of the local coefficient $c_2(h)=\tfrac12\phi''(h)$.
For Heaviside, saturated ReLU, and similar threshold–like gains, for $h<\theta$ the gain function is constant at zero output: $\phi(h)=0$ and hence $\phi'(h)=\phi''(h)=0$. 
Consequently $c_2\equiv 0$ and 
\begin{equation}
m(h,\sigma) \simeq c_1(h) \sigma - c_3(h) \sigma^3.
\end{equation}
For typical smooth sigmoids, $\phi'(x)$ is localized near the threshold and decays rapidly for $x\le\theta$.
In common exponential–tail cases, such as logistic and $\tanh$, this follows from their derivatives:
\emph{Logistic:}
\begin{equation}
\phi(x) = \frac{1}{1+e^{-x}}, \quad \phi'(x) = \frac{e^{-x}}{(1+e^{-x})^2},
\end{equation}
so for $x\le\theta$,
\begin{equation}
\phi'(x) \propto e^{-(\theta-x)},
\end{equation}
yielding exponential decay.

\emph{tanh:}
\begin{equation}
\phi(x) = \frac{1+\tanh(x)}{2}, \quad
\phi'(x) = \frac{1}{2}\big[1-\tanh^2(x)\big] \propto e^{-2|x|},
\end{equation}
so below threshold, $\phi'(x) \propto e^{-2(\theta-x)}$.

\emph{erf–sigmoid:} 
\begin{equation}
\phi(x) = \frac{1+\mathrm{erf}(x)}{2}, \quad \phi'(x) = \frac{1}{\sqrt{\pi}}  e^{-x^2},
\end{equation}
which decays super–exponentially as $e^{-{\rm const} (\theta-x)^2}$.
In these cases, differentiation preserves the exponential (or super–exponential) tail, so
\begin{equation}
|c_2(h)| = \tfrac12 |\phi''(h)|
\lesssim C  e^{-g\Delta} \quad\text{(logistic, $\tanh$)}, 
\end{equation}
or $|c_2(h)| \propto e^{-{\rm const} \Delta^2}$ for erf–sigmoid, with $\Delta=\theta-h\gg 1$.
For algebraic tails $\phi'(h)\sim \Delta^{-q}$ with $q>0$, one has $c_2\sim \Delta^{-(q+1)}$ and $c_3\sim \text{const}\cdot \Delta^{-3}$, hence
\begin{equation}
\frac{c_2^2}{c_3}  \sim  \Delta^{  1-2q}.
\end{equation}
If $q>\tfrac{1}{2}$, this ratio vanishes as $\Delta\to\infty$ and the cubic scaling holds uniformly.
By contrast, if the subthreshold slope does not decay, e.g.
\begin{equation}
\limsup_{\Delta\to\infty} \phi'(h)  >  0,
\end{equation}
then $c_2$ does not vanish whereas $c_1\sim \mu_0/(\pi\Delta)$, so $c_2/c_1$ does not go to zero and the quadratic term cannot be neglected.
Such gains lack a genuine off–regime and fall outside typical neural or machine–learning models.
Thus for standard gains (threshold–like, logistic, $\tanh$, erf), $|c_2|\ll c_1,c_3$ for large $\Delta$, so the quadratic term is irrelevant, while the leading monopole term $\mu_0>0$ guarantees $c_1>0$ and $c_3>0$ in the far field.
Inserting $c_1,c_3$ into the macroscopic dynamics
\begin{equation}
\tau\dot{\sigma}\simeq a(h) \sigma-b(h) \sigma^3,\quad a(h):=g c_1(h)-1,\quad b(h):=g c_3(h)>0,
\end{equation}
gives a continuous second-order transition with mean-field exponent $\beta=1/2$ and the same $t^{-1/2}$ critical slowing down at $a(h)=0$ as in the Heaviside case.

\subsection{Concrete examples}
We provide explicit coefficients for two standard activation models, showing their smooth convergence to the Heaviside limit.
\paragraph{Logistic sigmoid.}
For $\phi(x) = [1+\exp(-\gamma(x-\theta))]^{-1}$, the derivative is a localized bump $\phi' \propto \mathrm{sech}^2$.
The expansion coefficients take the form
\begin{equation}
    c_1(h) = \frac{1}{\pi\Delta}\left[1+\frac{\pi^2}{3 \gamma^2 \Delta^2} + \dots \right], \quad
    c_3(h) = \frac{1}{3\pi\Delta^3}\left[1 + \frac{2\pi^2}{\gamma^2 \Delta^2} + \dots \right],
    \label{eq:sigmoid_coefs}
\end{equation}
so that in the high-gain limit $\gamma \to \infty$ these recover the Heaviside coefficients $1/(\pi\Delta)$ and $1/(3\pi\Delta^3)$.
\paragraph{Saturated ReLU.}
For the piecewise linear function $\phi(x) = \min(1, \max(0, \gamma(x-\theta)))$, the derivative is a boxcar.
The coefficients are
\begin{equation}
    c_1(h) = \frac{1}{\pi\Delta}\left[1+\frac{1}{3 \gamma^2 \Delta^2} + \dots \right], \quad
    c_3(h) = \frac{1}{3\pi\Delta^3}\left[1 + \frac{2}{\gamma^2 \Delta^2} + \dots \right],
\end{equation}
again recovering the Heaviside limit as $\gamma \to \infty$.

\subsection{Effective and closed-loop sensitivities: general $A_{\mathrm{eff}}$ and saturation factor $K(\phi,m)$}
\label{subsec:general_Aeff_K}
The effective sensitivity is the $h$–derivative of the Poisson-smoothed rate $A_{\mathrm{eff}}(h,\sigma) \equiv \partial_h m(h,\sigma)$.
To obtain the near-onset scaling, we expand $A_{\mathrm{eff}}$ for small $\sigma$ using \eqref{eq:landau_expansion}:
\begin{equation}
A_{\mathrm{eff}}(h,\sigma) = \partial_h m(h,\sigma)
 = \phi'(h) - \sigma  \mathcal{H}(\phi'')(h) + \frac{\sigma^2}{2}\phi'''(h) + \frac{\sigma^3}{6}\mathcal{H}(\phi^{(4)})(h) + O(\sigma^4).
\label{eq:Aeff_expansion}
\end{equation}
In the subthreshold regime (where $\phi'(h)=0$ for threshold-like gains), the leading term is linear in $\sigma$, with coefficient set by the Hilbert transform of $\phi''$.
We therefore define the saturation factor $K(\phi,m)$ as the small–$\sigma$ slope of $A_{\mathrm{eff}}$:
\begin{equation}
K(\phi,m):=-\pi \mathcal{H}\big(\phi''\big)(h) = \pi  \partial_h c_1(h),
\label{eq:K_def}
\end{equation}
where $c_1(h)=-\mathcal{H}(\phi')(h)$ is the linear Landau coefficient.
Thus, for small $\sigma$,
\begin{equation}
A_{\mathrm{eff}}(h,\sigma) = \phi'(h) + \frac{K(\phi,m)}{\pi} \sigma + O(\sigma^2).
\label{eq:Aeff_approx}
\end{equation}
Using the multipole series for $c_1(h)$ \eqref{eq:c1expansion}, we obtain the moment series for $K(\phi,m)$:
\begin{equation}
K(\phi,m)=\sum_{k=0}^{\infty}(-1)^k(k+1)\frac{\mu_k}{\Delta^{k+2}}.
\label{eq:K_moment}
\end{equation}
The leading monopole term ($k=0$) is positive, ensuring $K(\phi,m)>0$ in the far field.
Near the critical point, expanding $m$ for small $\sigma$ gives
\begin{equation}
m(h,\sigma) \simeq c_1(h) \sigma - c_3(h) \sigma^3,\quad
\partial_\sigma m(h,\sigma) \simeq c_1(h) - 3c_3(h) \sigma^2.
\end{equation}
The stationary condition $\sigma_* = g  m(h,\sigma_*)$ yields
\begin{equation}
g c_3(h) \sigma_*^2 = a(h) := g c_1(h) - 1.
\end{equation}
The internal loop gain on the stationary manifold is
\begin{equation}
\mathcal{L}_*(h,\sigma) = g \partial_\sigma m(h,\sigma)\big|_{\sigma = \sigma_*}
\simeq g c_1(h) - 3a(h),
\end{equation}
so that $1-\mathcal{L}_* \simeq 2a(h)$.
Using \eqref{eq:Aeff_approx} with $\sigma=\sigma_*$ and $\phi'(h)=0$ (subthreshold), the closed-loop susceptibility \eqref{eq:dsigma_dh_appendix} admits the scaling
\begin{equation}
\frac{d\sigma_*}{dh}\simeq\frac{g  K(\phi,m) \sigma_*}{2\pi a(h)}
=\frac{K(\phi,m)}{2\pi c_3(h)}\frac{1}{\sigma_*}
\simeq\frac{K(\phi,m) c_1(h)}{2\pi c_3(h)}\frac{1}{m_*},
\label{eq:universal_scaling}
\end{equation}
where in the second equality we used $a(h)=g c_3(h)\sigma_*^2$ (from the fixed-point condition with $m\simeq c_1\sigma-c_3\sigma^3$), and in the last step $\sigma_*=g m_*$ with $g\simeq g_c=1/c_1(h)$ in the vicinity of the critical
point.
In the Heaviside limit, inserting $c_1=1/(\pi\Delta)$, $c_3=1/(3\pi\Delta^3)$ and $K=1/\Delta^2$ yields
\begin{equation}
    \frac{K(\phi,m) c_1}{2\pi c_3}=\frac{3}{2\pi},
\end{equation}
recovering the prefactor of the $1/m_*$ scaling of the closed-loop susceptibility found in the main text.

\section{Analytical Derivation of Cancellation Coefficients}
\label{sec:cancellation}
In the main text, we showed that when the threshold distribution follows a power law near the origin, the internal loop gain $\mathcal{L}_*(h,\sigma)$ and the effective input gain $g A_{\mathrm{eff}} (h,\sigma)$ become independent of the coupling strength $g$, and are instead fixed solely by the exponent $\kappa$ of the threshold statistics:
\begin{equation}
  \mathcal{L}_*(h,\sigma) = \kappa,\quad g A_{\mathrm{eff}}(h,\sigma) = \kappa\cot(\pi\kappa/2).
\end{equation}
Here we derive these cancellation coefficients explicitly.
We assume a general threshold density $\rho(\theta)$ that behaves asymptotically as a power law near the origin:
\begin{equation}
    \rho(\theta) \approx c_\rho  \theta^{\kappa-1} \quad (\text{as } \theta \to 0),
    \label{eq:power_law_density}
\end{equation}
where $c_\rho > 0$ is a scale-dependent constant and $0 < \kappa < 1$.
We evaluate the scalings of the population activity $m(\sigma)$ and the input sensitivity $A_{\text{eff}}(\sigma)$.
The population activity at $h=0$ is given by
\begin{equation}
    m(\sigma) = \frac{1}{\pi} \int_0^\infty \rho(\theta)  \arctan\left(\frac{\sigma}{\theta}\right) d\theta.
    \label{eq:activity_conv}
\end{equation}
Substituting the asymptotic form $\rho(\theta) \approx c_\rho \theta^{\kappa-1}$ and performing the change of variables $\theta = \sigma u$ yields
\begin{equation}
    m(\sigma) \approx \frac{c_\rho}{\pi} \sigma^\kappa \int_0^\infty u^{\kappa-1} \arctan\left(\frac{1}{u}\right) du = D  \sigma^\kappa, \quad D = \frac{c_\rho}{2\kappa} \sec\left(\frac{\pi\kappa}{2}\right).
    \label{eq:activity_scaling}
\end{equation}
The input sensitivity is
\begin{equation}
    A_{\text{eff}}(\sigma) = \partial_h m(h, \sigma)\big|_{h=0}
    = \frac{1}{\pi\sigma} \int_0^\infty \rho(\theta)  \frac{1}{1+(\theta/\sigma)^2} d\theta.
    \label{eq:Aeff_def_cancel}
\end{equation}
Using again $\theta = \sigma u$ and $\rho(\theta) \approx c_\rho \theta^{\kappa-1}$, we obtain
\begin{equation}
    A_{\text{eff}}(\sigma) \approx \frac{c_\rho}{\pi} \sigma^{\kappa-1}
    \int_0^\infty \frac{u^{\kappa-1}}{1+u^2} du = C  \sigma^{\kappa-1}, \quad C = \frac{c_\rho}{2} \csc\left(\frac{\pi\kappa}{2}\right).
    \label{eq:Aeff_scaling2}
\end{equation}

We now evaluate the two quantities that appear in the main text: the internal loop gain $\mathcal{L}_* = g \partial_\sigma m$ and the effective input gain $g A_{\mathrm{eff}}$.
\paragraph*{Internal loop gain $\mathcal{L}_*$.}
By definition \eqref{eq:def_L},
\begin{equation}
    \mathcal{L}_*(h, \sigma) =  g  \kappa  \frac{m(\sigma_*)}{\sigma_*} = \kappa,
\end{equation}
here we used $m(\sigma_*)/\sigma_* = 1/g$.
Thus, the stability margin $1-\mathcal{L}_*$ appearing in the curvature of the Lyapunov potential is fixed solely by $\kappa$, and is strictly positive for $0<\kappa<1$.
\paragraph*{Effective sensitivity $g A_{\mathrm{eff}}$.}
The effective sensitivity entering the closed-loop susceptibility is $g A_{\mathrm{eff}}(\sigma_*)$.
From \eqref{eq:Aeff_scaling2},
\begin{equation}
    g A_{\mathrm{eff}}(\sigma_*) = g  C  \sigma_*^{\kappa-1}.
\end{equation}
Using $\sigma_* = g D \sigma_*^\kappa$, we find
\begin{equation}
    g A_{\mathrm{eff}}(\sigma_*) = \left( \frac{\sigma_*^{1-\kappa}}{D} \right) C  \sigma_*^{\kappa-1}
    = \frac{C}{D}.
\end{equation}
Substituting the explicit forms of $C$ and $D$,
\begin{equation}
    \frac{C}{D} = \frac{(c_\rho/2)  \csc(\pi\kappa/2)}{(c_\rho/2\kappa)  \sec(\pi\kappa/2)}
    = \kappa \frac{\cos(\pi\kappa/2)}{\sin(\pi\kappa/2)}
    = \kappa \cot\left(\frac{\pi\kappa}{2}\right).
    \label{eq:gAeff_result}
\end{equation}
Both $\mathcal{L}_*$ and $g A_{\mathrm{eff}}$ thus exhibit a cancellation of $g$ and $c_\rho$, but they converge to different universal constants determined solely by the exponent $\kappa$.
\subsection{Specific Example: Gamma-Distributed Thresholds}
As a concrete check, consider the Gamma distribution
\begin{equation}
  \rho(\theta) = \frac{1}{\Gamma(\kappa)\Theta^\kappa} \theta^{\kappa-1} e^{-\theta/\Theta}.
\end{equation}
Near the origin, this matches the general form with $c_\rho = 1/[\Gamma(\kappa)\Theta^\kappa]$.
Substituting this specific $c_\rho$ yields the coefficients
\begin{equation}
    D_{\Gamma} = \frac{\sec(\pi\kappa/2)}{2\kappa \Gamma(\kappa)\Theta^\kappa}, \quad
    C_{\Gamma} = \frac{\csc(\pi\kappa/2)}{2 \Gamma(\kappa)\Theta^\kappa}.
\end{equation}
Their ratio is again $C_{\Gamma}/D_{\Gamma} = \kappa \cot(\pi\kappa/2)$, while $\mathcal{L}_*=\kappa$, confirming the general result.
\subsection{Marginal Case $\kappa=1$: Logarithmic Quasi-Cancellation}
\label{sec:marginal_case}
Finally, we address the marginal case $\kappa=1$, where the threshold density is finite at the origin.
We approximate the density for small thresholds by a constant with a soft cutoff at scale $\Theta$,
\begin{equation}
    \rho(\theta) \approx c_\rho \quad (\text{for } \theta \ll \Theta),
    \label{eq:const_density}
\end{equation}
where $c_\rho = \rho(0) > 0$.
Unlike the power-law case ($\kappa < 1$), the relevant integrals now diverge logarithmically, necessitating explicit inclusion of the cutoff $\Theta$.
To obtain the correct asymptotic behaviour of $\mathcal{L}_*$, we must retain sub-leading terms of order $O(1)$.
We first compute $\partial_\sigma m$.
Differentiating \eqref{eq:activity_conv} with respect to $\sigma$ and restricting the range to $[0, \Theta]$ gives
\begin{equation}
    \partial_\sigma m = \frac{1}{\pi} \int_0^\infty \rho(\theta)  \frac{\theta}{\sigma^2 + \theta^2} d\theta
    \approx \frac{c_\rho}{\pi} \int_0^{\Theta} \frac{\theta}{\sigma^2 + \theta^2} d\theta.
    \label{eq:chi_integral}
\end{equation}
Evaluating this integral yields
\begin{equation}
    \partial_\sigma m = \frac{c_\rho}{2\pi} \ln\left(1 + \frac{\Theta^2}{\sigma^2}\right)
    \approx \frac{c_\rho}{\pi} \ln\left(\frac{\Theta}{\sigma}\right).
    \label{eq:chi_log_result}
\end{equation}
The population activity is obtained by integrating with respect to $\sigma$, keeping the same order of approximation.
Using $\int \ln(A/x)  dx = x \ln(A/x) + x$, we find
\begin{equation}
    m(\sigma) = \int_0^\sigma \partial_\sigma m  ds
    \approx \frac{c_\rho}{\pi} \int_0^\sigma \ln\left(\frac{\Theta}{s}\right) ds
    = \frac{c_\rho}{\pi} \sigma \left[ \ln\left(\frac{\Theta}{\sigma}\right) + 1 \right].
    \label{eq:y_log_result}
\end{equation}
The stationary point is determined by $\sigma_* = g  m(\sigma_*)$, so
\begin{equation}
    \sigma_* = g \frac{c_\rho}{\pi} \sigma_* \left[ \ln\left(\frac{\Theta}{\sigma_*}\right) + 1 \right]
    \implies g \frac{c_\rho}{\pi} = \frac{1}{\ln(\Theta/\sigma_*) + 1}.
    \label{eq:gamma_log_cond}
\end{equation}
The internal loop gain at this point is
\begin{equation}
    \mathcal{L}_*(\sigma_*) = g  \partial_\sigma m
    = g \frac{c_\rho}{\pi} \ln\left(\frac{\Theta}{\sigma_*}\right).
    \label{eq:L_log_sub}
\end{equation}
Substituting \eqref{eq:gamma_log_cond}, we obtain
\begin{equation}
    1-\mathcal{L}_*(h,\sigma) =
    1-\frac{\ln(\Theta/\sigma_*)}{\ln(\Theta/\sigma_*) + 1}
    = 1-\frac{1}{1 + \frac{1}{\ln(\Theta/\sigma_*)}}
    \approx \frac{1}{\ln(\Theta/\sigma_*)}.
    \label{eq:L_log_approx}
\end{equation}
So that for $\kappa=1$ the system is no longer pinned to a constant negative eigenvalue, but instead approaches a quasi-marginal regime as $\sigma_* \to 0$ (or $g \to \infty$).
For the exponential distribution $\rho(\theta) = \Theta^{-1}e^{-\theta/\Theta}$ (Gamma with $\kappa=1$), we have $c_\rho = \rho(0) = 1/\Theta$, and the logarithmic coefficient is $D_{\log} = c_\rho/\pi = 1/(\pi\Theta)$.

\section{Kernel–agnostic bounds for sign–changing kernels and $1<\alpha<2$}
\label{sec:alpha_bounds}
In the main text, we showed that for stability index $\alpha = 1$ (Cauchy case) and nonnegative kernels $G(t)\ge 0$, the absolute value inside the convolution can be dropped, leading to a one–dimensional macroscopic closure.
For general symmetric $\alpha$–stable laws with $1<\alpha<2$, or for sign–changing kernels, this simplification no longer holds.
In this section we derive kernel–agnostic inequalities that bound the macroscopic noise scale from above and below.
These bounds hold for arbitrary sign structure of $G$ and for all $1<\alpha\le 2$, and indicate that the key ingredient of the AGC mechanism—activity–dependent amplification of the noise scale and the resulting suppression of effective sensitivity—remains robust beyond the solvable Cauchy case.
We consider a causal kernel $G\in L^1(\mathbb{R}_+)$, a bounded non–decreasing rate function $\phi:\mathbb{R}\to\mathbb{R}$ (normalized to $0\le \phi\le 1$ in the main text), and define
\begin{equation}
\|G\|_1:=\int_0^\infty |G(s)|ds,\quad K_0:=\int_0^\infty G(s) ds .
\end{equation}
For a symmetric $\alpha$–stable law $S_\alpha(\sigma)$, the characteristic function is
\begin{equation}
\langle e^{i\lambda X} \rangle = \exp \left(-\sigma^\alpha |\lambda|^\alpha\right).
\end{equation}
Comparing with \eqref{eq:log_char_functional}, the factor multiplying $-|\lambda|^\alpha$ is naturally identified with $\sigma_\alpha(t)^\alpha$.
We therefore define
\begin{equation}
\sigma_\alpha(t):= g^{1/\alpha}\left\langle \left| \int_0^t G(t-s)  \phi(x(s))  ds \right|^\alpha \right\rangle^{1/\alpha}_{P[x]},
\end{equation}
so that the filtered input $\tilde{x}(t)$ has $\alpha$–stable scale $\sigma_\alpha(t)$ at each time $t$.

\subsection{Two–sided bounds on $\sigma_\alpha(t)$}
We first derive general upper and lower bounds for $\sigma_\alpha(t)$ that hold for sign–changing kernels $G$ and bounded $\phi$.
For $1<\alpha\le 2$, the function $z\mapsto |z|^\alpha$ is convex, so Jensen's inequality gives
\begin{equation}
\left\langle \Big|\int_0^t G(t-s) \phi(x(s))ds\Big|^\alpha\right\rangle
\ \ge\ \Big|\left\langle \int_0^t G(t-s) \phi(x(s))ds\right\rangle\Big|^\alpha .
\end{equation}
Using Fubini's theorem, we can interchange the expectation and the time integral:
\begin{equation}
\left\langle \int_0^t G(t-s) \phi(x(s))ds \right\rangle
= \int_0^t G(t-s) \langle \phi(x(s))\rangle  ds
= (G * m)(t).
\end{equation} 
This yields the general lower bound
\begin{equation}
\sigma_\alpha(t)\ \ge\ g^{1/\alpha} \big|(G\ast m)(t)\big|.
\label{eq:LowerBoundGeneralAlpha}
\end{equation}
For the upper bound, we start from the triangle inequality:
\begin{equation}
\left| \int_0^t G(t-s)  \phi(x(s))  ds \right|
\le\int_0^t |G(t-s)| \phi(x(s)) ds .
\end{equation} 
Define $a(s) := |G(t-s)| \ge 0$ and $b(s) := \phi(x(s)) \in [0, M]$, where $M := \|\phi\|_\infty$ is the maximum of $\phi$.
Applying Hölder's inequality in the form
\begin{equation}
\left( \int_0^t a(s)  b(s) ds \right)^\alpha
\le\left( \int_0^t a(s) ds \right)^{\alpha-1} \int_0^t a(s) b(s)^\alpha ds
\end{equation} 
and using $b^\alpha \le M^{\alpha-1} b$ gives
\begin{equation}
\left( \int_0^t a b \right)^\alpha
\le\|G\|_1^{\alpha-1} M^{\alpha-1} \int_0^t |G(t-s)| \phi(x(s)) ds .
\end{equation} 
Taking the expectation over $x(\cdot)$ and noting that
\begin{equation}
\left\langle \int_0^t |G(t-s)|  \phi(x(s)) ds \right\rangle
= (|G| * m)(t),
\end{equation} 
we obtain the general upper bound
\begin{equation}
\sigma_\alpha(t)\ \le\ g^{1/\alpha}  M^{1 - 1/\alpha} \|G\|_1^{1 - 1/\alpha} 
\big[ (|G| * m)(t) \big]^{1/\alpha}.
\label{eq:UpperBoundGeneralAlpha_rewrite}
\end{equation}
In a stationary regime with constant activity $m(t) \equiv m_*$, the convolutions simplify to
$(G * m_*)(t) = K_0  m_*$ and $(|G| * m_*)(t) = \|G\|_1  m_*$.
Applying \eqref{eq:LowerBoundGeneralAlpha}–\eqref{eq:UpperBoundGeneralAlpha_rewrite}
then yields
\begin{equation}
g^{1/\alpha}|K_0| m_* \ \le\ \sigma_{\alpha*}\ \le\ g^{1/\alpha}M^{1-1/\alpha} \|G\|_1 m_*^{1/\alpha}.
\label{eq:StationaryTwoSided}
\end{equation}
For the normalization $0\le \phi\le 1$ and $\|G\|_1=1$ used in the main text, this simplifies to
\begin{equation}
g^{1/\alpha}|K_0|  m_* \ \le\ \sigma_{\alpha*}\ \le\ g^{1/\alpha}  m_*^{1/\alpha}.
\end{equation}
\subsection{Effective sensitivity bounds}
To connect these bounds to the robustness of critical dynamics, we next control the effective sensitivity $A_{\mathrm{eff}}(h,\sigma_\alpha)$ as a function of the noise scale $\sigma_\alpha$.
Let $p_{\alpha,\sigma}(u) = \sigma^{-1} p_{\alpha,1}(u/\sigma)$ denote the probability density function of a symmetric $\alpha$-stable ($S\alpha S$) distribution with scale $\sigma$. Its peak density scales inversely with the width:
\begin{equation}
    \sup_{x} p_{\alpha,\sigma}(x) = \frac{C_\alpha}{\sigma}, \quad
    C_\alpha = \frac{\Gamma(1/\alpha)}{\pi \alpha},
\end{equation}
so that $C_1=1/\pi$ (Cauchy).
The effective sensitivity is defined as the susceptibility of the population rate to changes in the mean input $h$:
\begin{equation}
    A_{\mathrm{eff}}(h,\sigma_\alpha) := \partial_h m(h,\sigma_\alpha)
    = \partial_h \int_{\mathbb{R}} p_{\alpha,\sigma_\alpha}(x-h) \phi(x) dx.
\end{equation}
Intuitively, larger noise $\sigma_\alpha$ smooths out the activation function $\phi(x)$, reducing the slope of the population response.
Mathematically, we use a simple regularity assumption on $\phi$. We say that the rate function $\phi$ has \emph{bounded variation} if the total amount by which it increases and decreases over $\mathbb{R}$ is finite.
All monotone gain functions with range $[0,1]$ (Heaviside, sigmoids, saturated ReLUs) are of this type.
Since $\phi$ is bounded and non–decreasing, it induces a finite Lebesgue–Stieltjes measure on $\mathbb{R}$, which we denote by $d\phi(x)$.
More precisely, for any bounded continuous test function $f$, we define $\int f(x) d\phi(x)$ as the Stieltjes integral with respect to the measure $\mu_\phi$ given by $\mu_\phi((a,b])=\phi(b)-\phi(a)$.
If $\phi$ is absolutely continuous, then $d\phi(x)=\phi'(x) dx$, while
jump discontinuities of $\phi$ contribute atomic masses.
For such $\phi$, the convolution can be written as a Stieltjes integral with respect to $\phi$:
\begin{equation}
    m(h,\sigma_\alpha)
    = \int_{\mathbb{R}} \phi(x) p_{\alpha,\sigma_\alpha}(x-h) dx
    = \int_{\mathbb{R}} p_{\alpha,\sigma_\alpha}(x-h) d\phi(x),
\end{equation}
so that
\begin{equation}
    A_{\mathrm{eff}}(h,\sigma_\alpha)
    = \partial_h m(h,\sigma_\alpha)
    = \int_{\mathbb{R}} \partial_h p_{\alpha,\sigma_\alpha}(x-h) d\phi(x)
    = - \int_{\mathbb{R}} \partial_x p_{\alpha,\sigma_\alpha}(x-h) d\phi(x).
\end{equation}
Taking absolute values and using the standard bound for Stieltjes
integrals,
\begin{equation}
  \bigg|\int f(x) d\phi(x)\bigg|
  \le \sup_x |f(x)| \|\phi\|_{\mathrm{TV}},
\end{equation}
we obtain
\begin{equation}
    |A_{\mathrm{eff}}(h,\sigma_\alpha)|
    \le \sup_x |\partial_x p_{\alpha,\sigma_\alpha}(x)| \|\phi\|_{\mathrm{TV}}.
\end{equation}
Here $\|\phi\|_{\mathrm{TV}}$ denotes the total variation of $\phi$, i.e.
\begin{equation}
\|\phi\|_{\mathrm{TV}}
= \sup_{P}\sum_j |\phi(x_j)-\phi(x_{j-1})|,
\end{equation}
with the supremum over all finite partitions $P$ of $\mathbb{R}$.
The derivative of the $\alpha$-stable density is bounded, and its supremum scales as $1/\sigma_\alpha$ up to an $\alpha$-dependent constant, so we can write
\begin{equation}
    |A_{\mathrm{eff}}(h,\sigma_\alpha)|
    \le \frac{C_\alpha}{\sigma_\alpha} \|\phi\|_{\mathrm{TV}}.
\end{equation}
Thus we obtain the bound
\begin{equation}
    A_{\mathrm{eff}}(h,\sigma_\alpha)
    \le \frac{C_\alpha}{\sigma_\alpha} \|\phi\|_{\mathrm{TV}}.
    \label{eq:AeffBoundVariation}
\end{equation}
For the class of monotone gain functions with range $[0,1]$ used in the main text (Heaviside, sigmoids, saturated ReLUs), we have $\|\phi\|_{\mathrm{TV}} = \phi(+\infty)-\phi(-\infty) = 1$, so the bound simplifies to
\begin{equation}
    A_{\mathrm{eff}}(h,\sigma_\alpha)
    \le \frac{C_\alpha}{\sigma_\alpha}.
    \label{eq:AeffSimple}
\end{equation}
In other words, increasing the $\alpha$-stable noise scale $\sigma_\alpha$ suppresses the effective sensitivity of the population response at least as fast as $1/\sigma_\alpha$, uniformly over all monotone gains considered here.

\subsection{Kernel-agnostic AGC bound}
We now combine the sensitivity bound \eqref{eq:AeffSimple} with the lower bound on the noise scale from the previous subsection.
For networks where the synaptic kernel has a non-zero mean (e.g.\ predominantly excitatory or inhibitory kernels), the stationary bound \eqref{eq:StationaryTwoSided} gives $\sigma_{\alpha*} \ge g^{1/\alpha} |K_0| m_*$.
Substituting this into \eqref{eq:AeffSimple} yields
\begin{equation}
    A_{\mathrm{eff}}(h,\sigma_{\alpha*}) \le \frac{C_\alpha}{\sigma_{\alpha*}}
    \le \frac{C_\alpha}{g^{1/\alpha} |K_0|  m_*}.
    \label{eq:AGCcertificate}
\end{equation}
This inequality provides a kernel-agnostic upper bound on the effective sensitivity: as the population activity $m_*$ increases, the effective sensitivity necessarily decays at least as fast as $1/m_*$.
This implements a divisive-like automatic gain control: large activity levels generate large input fluctuations, which in turn smooth the effective transfer function and suppress further amplification, preventing runaway excitation independently of the detailed shape of $G(t)$.

\section{Details of the simulations}
All simulations were implemented in Python using standard numerical libraries: NumPy, SciPy, Numba, Matplotlib.
We simulate sparse rate networks of $N=8000$ units for Fig.~1 and $N=6000$ for Fig.~2 with Heaviside gain $\phi(x)=\Theta(x-\theta)$ and threshold $\theta=1$.
Each neuron receives $K_{\rm in}$ random presynaptic inputs from the remaining $N-1$ neurons (no self–couplings, no multiple edges), with $K_{\rm in}=256$ for Fig.~1 and $K_{\rm in}=128$ for Fig.~2.
Synaptic weights are independent Cauchy variables $J_{ij}=(g/K_{\rm in})X_{ij}$, $X_{ij}\sim{\rm Cauchy}(0,1)$.
The dynamics follow
\begin{equation}
  \tau \dot x_i(t) = -x_i(t) + \eta_i(t) + h,\quad
  \eta_i(t) = \sum_{j\neq i} J_{ij} \Theta(x_j(t)-\theta),
\end{equation}
with $\tau=1$ and homogeneous input $h$.
We integrate with a forward Euler scheme of step $\Delta t$ and use independent seeds for different realizations.
\subsection*{Fig.~1 and Fig.~S1}
For the left panel of Fig.~1 and for Fig.~S1 we set $h=0.8$, $\Delta t=0.002$ and run $T_{\rm sim}=8000$ steps for $g\in\{1.0,1.2,1.4\}$ and $n_{\rm seeds}=20$.
Initial conditions are drawn from $z_i(0)\sim{\rm Cauchy}(0,\sigma_0)$ with $\sigma_0=0.05$, and $x_i(0)=z_i(0)+h$.
We introduce auxiliary filtered inputs $z_i^{(L)}$ obeying
\begin{equation}
  \tau \dot z_i^{(L)}(t) = -z_i^{(L)}(t) + L \eta_i(t),
\end{equation}
corresponding to exponential kernels $G_L(t)=(L/\tau)e^{-t/\tau}\Theta(t)$ with
$L\in\{0.5,1.0,1.5,2.0\}$ and $\|G_L\|_1=L$.
The macroscopic scale is estimated from a random subset of $M=6000$ neurons as $\hat\sigma(t)=\tfrac12(Q_{75}-Q_{25})$ of $\{z_i^{(1)}(t)\}$, where $Q_{p}$ denotes the $p$-th percentile (interquartile range, IQR).
Stationary values $\hat\sigma_{\rm ss}^{(L)}$ and $m_*$ are extracted from the second half of the simulation and averaged across seeds.
The mean-field ODE for $\sigma(t)$ is solved with the same $\Delta t$ (fourth–order Runge–Kutta) from $\sigma(0)=\hat\sigma(0)$ and compared to $\hat\sigma(t)$ in the left panel of Fig.~1; Fig.~S1 shows Q–Q plots of standardized $z_i^{(1)}(t)$ and the $L^1$ scaling of $\hat\sigma_{\rm ss}^{(L)}$.
The center and right panels of Fig.~1 probe the phase transition and critical relaxation.
Here we keep the same architecture but set $h=0.6$ and $\Delta t=10^{-2}$.
The theoretical critical coupling is $g_c^{\rm th}=\pi(\theta-h)$.
For the center panel we use a fixed connectivity $(J_0,{\rm pre\_idx})$ and scan $g\in g_c^{\rm th}\times\{1.01,1.02,1.04,1.06,1.08,1.10\}$, with $n_{\rm trials}=20$ per $g$.
For each trial we perform a burn-in of $T_{\rm burn}=10^4$ steps, including an initial boost phase of length $T_{\rm seed}=600$ with input $h+m_{\rm boost}$, $m_{\rm boost}=0.5$, followed by a measurement window of $T_{\rm meas}=10^4$ steps at input $h$.
We estimate $m_*$ and $\sigma_*$ from the second half of the window ($\hat\sigma$ as IQR/2 over all $N$ units), compute the mean and standard error of $\sigma_*^2$ across trials, and fit $\sigma_*^2 \approx S_{\rm iqr} g + I_{\rm iqr}$.
The empirical critical point is given by $g_c^{\rm exp}=-I_{\rm iqr}/S_{\rm iqr}$; the ratio $\sigma_*/(g m_*)$ is also monitored to check $\sigma_*=g m_*$.
The right panel of Fig.~1 probes the critical relaxation at
$g^\star=g_c^{\rm th}$.
We perform $n_{\rm trials}20$ decay experiments: each run is driven with $h+m_{\rm boost}$ for $T_{\rm seed}$ steps to reach a high-activity state, then relaxed at input $h$ for $T_{\rm meas}=10^4$ steps.
We record $\sigma(t)$ (IQR/2 of $z_i(t)$) and $m(t)$ at each step and average over trials.
On the window $t\in[T_{\min},T_{\max}]$ with $T_{\min}=10$, $T_{\max}=30$ (in units of $\tau$) we fit the slopes of $\log\sigma(t)$ and $\log m(t)$ versus $\log t$.
The right panel of Fig.~1 shows the mean $\pm$ SEM together with a $t^{-1/2}$ guide; we also check that the corresponding macroscopic ODE for $\sigma(t)$ exhibits the same decay exponent.
\subsection*{Fig.~2}
For Fig.~2 we study near-critical gain control and kernel–agnostic
bounds.
We use $N=6000$, $K_{\rm in}=128$, $\theta=1$, baseline input $h=m_{\rm base}=0.8$, $\tau_x=\tau_{\rm syn}=1$, $\Delta t=10^{-2}$, and a burn-in of $T_{\rm burn}=6000$ steps.
In all runs we estimate $m_*$ as the time average and $\sigma_*$ as the time-median of the spatial IQR/2 of the filtered inputs over the last $40\%$ of a DC window of length $T_{\rm dc}=10^4$, and then average across trials.
The left and center panels of Fig.~2 use purely exponential kernels $G(t)=\tau_{\rm syn}^{-1}e^{-t/\tau_{\rm syn}}\Theta(t)$.
We scan $g\in\{0.67, 0.68, 0.69, 0.72, 0.74, 0.76, 0.78, 0.80, 0.9, 1.0, 1.2, 1.4, 1.7, 2.0\}$ with $n_{\rm trials}=30$ each, measure $(m_*,\sigma_*)$, and verify $\sigma_*/(g m_*)\approx 1$.
To obtain $d\sigma_*/dh$ we repeat the DC protocol from the same post–burn-in state for six nearby inputs $h=m_{\rm base}\pm k\varepsilon$, $k\in\{1,2,3\}$, with $\varepsilon$ chosen adaptively to target $\Delta\sigma\approx 0.015$ and bounded in $[10^{-4},10^{-2}]$.
A least-squares fit of $\sigma_*$ versus $h$ gives $d\sigma_*/dh$, which we use with
\begin{equation}
  1-\mathcal{L}_*(m_*)
  = \frac{\sin^2(\pi m_*)}{\pi m_* d\sigma_*/dh}
\end{equation}
to reconstruct the loop gain deficit in the left panel; the center panel plots $d\sigma_*/dh$ versus $m_*$ on log–log axes and compares to the exact Heaviside prediction and its asymptotic $3/(2\pi m_*)$ behavior.
The right panel of Fig.~2 tests sandwich bounds for biphasic kernels
\begin{equation}
  G(t) = a h_{\tau_1}(t) - b h_{\tau_2}(t),\quad
  h_{\tau}(t)=\tau^{-1}e^{-t/\tau}\Theta(t),
\end{equation}
with $a=1$, $\tau_1=1$, $\tau_2=0.3$, and $b\in\{0.8,1.2,1.8\}$.
For each $(g,b)$ with $g\in\{1.6,1.8,2.0,2.2,2.4\}$ and $20$ trials we simulate two auxiliary filters $u_i$ and $v_i$ for $h_{\tau_1}$ and $h_{\tau_2}$ and define $z_i(t)=a u_i(t)-b v_i(t)$.
From the tail window we estimate $m_*$ and $\sigma_*$.
We compute
\begin{equation}
  K_0 = \int_0^\infty G(t) dt = a-b,\quad
  \|G\|_1 = \int_0^\infty |G(t)| dt,
\end{equation}
where $\|G\|_1$ is evaluated analytically in closed form, and define
\begin{equation}
  \sigma_{\rm lower} = g m_* |K_0|,\quad
  \sigma_{\rm upper} = g m_* \|G\|_1.
\end{equation}
\end{document}